\documentclass[]{interact}

\usepackage{epstopdf}
\usepackage[caption=false]{subfig}

\usepackage{amsmath}
\usepackage{url}
\usepackage{verbatim}
\usepackage{fancyvrb}
\usepackage{xcolor}
\usepackage[numbers,sort&compress]{natbib}
\bibpunct[, ]{[}{]}{,}{n}{,}{,}
\renewcommand\bibfont{\fontsize{10}{12}\selectfont}
\makeatletter
\def\NAT@def@citea{\def\@citea{\NAT@separator}}
\makeatother

\theoremstyle{plain}

\theoremstyle{definition}

\theoremstyle{remark}

\newcommand{\ed}{\end{document}}

\newcommand{\vy}{{\bm y}}

\newcommand{\vmu}{{\bm \mu}}
\newcommand{\veta}{{\bm \eta}}
\newcommand{\ca}[1]{\ensuremath{[#1]_h}}
\newcommand{\uno}[1]{\ensuremath{\left|#1\right|}}

\DefineVerbatimEnvironment{Code}{Verbatim}{fontshape=sl}

\begin{document}

\articletype{ARTICLE TEMPLATE.  Words' count: 8462}

\title{\texttt{pivmet}: Pivotal Methods for Bayesian Relabelling and $k$-Means Clustering}

\author{
\name{Leonardo Egidi\textsuperscript{a}\thanks{CONTACT L. Egidi. Email: legidi@units.it}, Roberta Pappad\`{a}\textsuperscript{a}, Francesco Pauli \textsuperscript{a} and Nicola Torelli \textsuperscript{a}}
\affil{\textsuperscript{a} Department of Economics, Business, Mathematics and Statistics, University of Trieste, Via Tigor 22, 34124, Trieste, Italy}
}

\maketitle

\begin{abstract}
The identification of groups' prototypes, i.e. elements of a dataset that represent different groups of data points, may be relevant to the tasks of clustering, classification and mixture modeling. 
The \texttt{R} package \texttt{pivmet} presented in this paper includes different methods for extracting pivotal units from a dataset. One of the main applications of pivotal methods is a Markov Chain Monte Carlo (MCMC) relabelling procedure to solve the label switching in Bayesian estimation of mixture models. 
 Each method returns posterior estimates, and a set of graphical tools for visualizing the output. The package offers JAGS and Stan sampling procedures for Gaussian mixtures, and allows for user-defined priors' parameters. 
The package also provides functions to perform consensus clustering based on pivotal units, which may allow to improve classical techniques (e.g. $k$-means) by means of a careful seeding. The paper provides examples of applications to both real and simulated datasets. 
\end{abstract}

\begin{keywords}
Pivotal units; mixture models; \texttt{pivmet}; relabelling; consensus clustering
\end{keywords}

\section[Introduction]{Introduction} 
\label{sec:intro}

The identification of some units which may be representative of the group they belong to is often a matter of statistical importance. In the so called big-data age, summarizing some essential information from a data pattern is often relevant and can help avoiding an extra amount of work when processing the data. The advantage of such pivotal units (hereafter called pivots) is that they are somehow chosen to be  as far as possible from units in the other groups and as similar as possible to the units in the same group. Despite the lack of a strict theoretical framework behind their characterization, the pivots may be beneficial in many statistical frameworks, such as clustering, classification, and mixture modeling. In this paper we will mainly focus on the latter, by showing how the use of pivotal quantities may provide a valid tool to derive reliable estimates.  

Under a Bayesian perspective, the estimation of mixture models suffers from the nonidentifiability of the mixture parameters during the MCMC sampling, a phenomenon known as `label switching'.  Because the likelihood is invariant under any permutation of the groups' labels \cite{stephens2000dealing}, then the inferences on component specific parameters are very poor. Some identifiability constraints have been proposed by \cite{richardson1997bayesian} and \cite{fruhwirth2001markov}, however in some circumstances it can be difficult to assume an ordering, especially in high dimensions. Finding suitable identifiability conditions might also be an issue in low dimensions if some clusters vary with respect to the means and others with respect to the variances.
 
The most popular strategy to cope with label switching is to use relabelling, a class of algorithms designed to permute the Markov chains in such a way to obtain reliable inferences  \cite{jasra2005markov}. The {\tt R} package {\tt label.switching} \cite{ papastamoulis2016label, label.switching} includes the methods in \cite{marin2007bayesian, marin2005bayesian, papastamoulis2010artificial, stephens2000dealing, rodriguez2014label, sperrin2010probabilistic}. The functions implemented in the package can be used to obtain the relabelled posterior estimates for the means, the standard deviations and the mixture weights of the fitted model. Some of these methods may be demanding in terms of computational complexity and require the full chains approximating the posterior distribution. 

The {\tt pivmet} package \cite{pivmet} for \texttt{R}, available from the Comprehensive \texttt{R} Archive Network at \url{http://CRAN.R-project.org/package=pivmet}, implements various pivotal selection criteria and the relabelling method described in \cite{egidi2018relabelling}. Graphical tools are also provided, in order to detect the label switching phenomenon and to check the method effectiveness. Moreover, compared to other packages whose architecture is focused on just one computational method, the user may fit its own mixture model either via the JAGS \cite{rjags} or the Stan \cite{rstan}  software, enjoying a non-negligible extent of flexibility to elicit prior distributions. The core of the implemented methodology is to detect pivotal units via the similarity matrix derived from the MCMC sample, whose elements are the probabilities that any two units in the observed sample are drawn from the same component; then, the pivots are used to relabel the chains.

Pivotal units may be fruitfully used in Dirichlet process mixtures (DPMM) \cite{ferguson1983bayesian, escobar1995bayesian, neal2000markov}, a class of models that naturally sorts data into clusters. An illustration will be proposed in the simulation section.

Another context where pivots identification may be beneficial is data clustering, as discussed in \cite{kmeans}. A careful seeding based on well-separated statistical units can be used to overcome  well-known limitations of the standard  $k$-means algorithm  and improve over the final clustering solution, especially in the case of imbalanced groups. The {\tt pivmet} package allows to implement a variant of the $k$-means algorithm, where the pivots are chosen via consensus clustering.

The rest of the paper is organized as follows. Section~\ref{sec:models} reviews mixture models, the relabelling procedure and the pivotal methods proposed by \cite{egidi2018relabelling}. The main functions of the \texttt{R} package \texttt{pivmet} are introduced in Section~\ref{sec:pivmet} along with some simulated examples, whereas two real data examples are presented in Section~\ref{sec:illustrations}. Section~\ref{sec:summary} concludes.

\section{Mixture models and label switching} 
\label{sec:models}

Let $\vy=(y_1,\ldots,y_n)$ denote a sample of $n$ observations; assume that $\bm{z}=(z_1,\ldots,z_n)$ is an unobserved latent sequence of independent and identically distributed random variables, with each $z_i \in \{1,\ldots,k\}$, where $k>1$ is a known integer denoting the number of components. Conditionally to $z_i$, a component-specific vector parameter $\vmu=(\mu_{1},\dots,\mu_{k})$, and a parameter $\phi$  which is common to all components, we have:
\begin{equation}
(Y_i|z_i=j, \mu_j, \phi) \sim f(y_i;\mu_j,\phi),  \ \ \ j=1,\dots,k
\label{eq:ydist} 
\end{equation}
for $i=1, \dots, n$, where $f(\cdot; \mu_j,\phi)$ denotes a parametric distribution.

Assume that $z_1,\ldots,z_n$ follow the multinomial distribution with weights  $\veta=(\eta_{1},\dots,\eta_{k})$, such that:
\begin{equation}
	P(z_i=j)=\eta_j\label{eq:gdist}, \ \ \ j=1,\dots,k.
\end{equation}
The likelihood of the mixture model is then
\begin{equation}
L(\vy;\vmu,\veta,\phi) = \prod_{i=1}^n \sum_{j=1}^k \eta_j f(y_i;\mu_j,\phi).
\label{eq:lik}
\end{equation}
Let $\mathcal{V}$ be the set of all the permutations of $\{1,\ldots,k \}$ and let $\nu \in  \mathcal{V}$. Let $\nu(\bm{\mu})= (\mu_{\nu(1)},\ldots,\mu_{\nu(k)})$ and $ \nu(\bm{\eta})=(\eta_{\nu(1)},\ldots,\eta_{\nu(k)})$ be the corresponding permutations of $\vmu$ and $\veta$, respectively. The likelihood in \eqref{eq:lik} is invariant under any permutation $\nu \in \mathcal{V}$, that is
\begin{equation}
L(\vy;\vmu,\veta,\phi) = L(\vy;\nu(\bm{\mu}),\nu(\bm{\eta}),\phi). \label{eq:invlik}
\end{equation}
As a consequence, the model is unidentified with respect to an arbitrary permutation of the labels. The label switching phenomenon occurs when performing Bayesian inference for the model \eqref{eq:lik}. If the prior distribution $p_0(\vmu,\veta,\phi)$ is invariant under a permutation of the indexes, that is 
\begin{equation}
p_0(\vmu,\veta,\phi) = p_0(\nu(\bm{\mu}),\nu(\bm{\eta}),\phi),
\end{equation}
then the posterior distribution of the mixture model parameters expressed as $p(\vmu,\veta,\phi|\vy) \propto$  $p_0(\vmu,\veta,\phi)$ $L(\vy;\vmu,\veta,\phi)$ is multimodal with (at least) $k!$ modes. It follows that the same invariance property holds for the posterior distribution, that is, 
\begin{equation}
p(\nu(\vmu),\nu(\veta),\phi|\vy)=p(\vmu, \veta, \phi|\vy)
\end{equation}
for all $\nu$, $\vmu$, $\veta$, $\phi$. This implies that  all simulated parameters should be switched to one among the $k!$ symmetric areas of the posterior distribution, by applying suitable permutations of the labels to each MCMC draw.

As mentioned in the Introduction, most of the existing approaches to perform inferences in the presence of label switching are based on the relabelling of the MCMC chain (see \cite{papastamoulis2016label} for a recent  and comprehensive review).  Note that the relabelling issue becomes relevant when we are interested, directly or indirectly, in the features of $k$ groups, such as the
posterior (and predictive) distributions of component-related quantities (e.g., the probability of each unit belonging to each group). In general, relabelling strategies may act during the MCMC sampling, and/or they may be used to post-process the chains, which  can be particularly convenient since the issue can be ignored in performing the MCMC.

The idea of solving the relabelling issue by fixing the groups for some units was firstly investigated by \cite{chung2004difficulties}, where, however, no indication on how to choose the units was supplied.
In \cite{marin2005bayesian, marin2007bayesian} the Pivotal Reordering Algorithm ({\tt PRA}) is proposed, which is based on a permutation of all simulated MCMC samples of parameters so that they are maximizing their similarity to a pivot parameter vector. Suitable pivotal quantities are also defined in the {\tt ECR} algorithm by \cite{papastamoulis2010artificial} via the equivalence classes representatives, and in \cite{yao2014online}, who propose relabelling each iteration by minimizing some distance from a reference labelling.

In the following Section we will summarize the relabelling procedure introduced in \cite{egidi2018relabelling} that is provided by the package {\tt pivmet}. The main idea is to use the \emph{pivots}, which are (pairwise) separated units with (posterior) probability one, in order to perform the post-MCMC relabelling of the chains.

\subsection{Pivotal relabelling}
\label{sec:relabelling}

We assume that an MCMC sample is obtained from the posterior distribution for model (\ref{eq:lik}) with a prior distribution which is labelling invariant.
We denote as $\{\ca{\theta}:h=1,\ldots,H\}$ the sample for the parameter $\theta=(\vmu,\veta,\phi)$, $H$ being the number of MCMC iterations.
We assume that a MCMC sample for the variable $\bm{z}$ is also obtained and we denote it by  $\{\ca{\bm{z}}:h=1,\ldots,H\}$.

Starting from a \emph{reference} partition 
\begin{equation}\mathcal{P}=\{\mathcal K_1,\ldots,\mathcal K_{{k}}\}
\label{eq:partition}
\end{equation}
of the $y_i$'s into $k$ non-overlapping groups, we define the probability of two units being in the same group across the MCMC sample as $c_{ip}=P(z_i=z_p|\vy)$, for $i=1,\ldots,n$ and $p=1,\ldots,n$. The estimate of $c_{ip}$ based on the MCMC sample is
\begin{equation}
\hat{c}_{ip} = \frac{1}{H} \sum_{h=1}^H \uno{\ca{z_i}=\ca{z_p}},
\label{eq:Cmatrix}
\end{equation}
where $|\cdot|$ denotes the indicator function of an event.
The $n\times n$ matrix $\hat{C}$ with elements $\hat{c}_{ip}$ can be seen as an estimated similarity matrix between units, and the complement to one $\hat{s}_{ip}=1-\hat{c}_{ip}$ as a dissimilarity matrix (note that $s_{ip}=0$ does not imply that the units $i$ and $p$ are the same, and therefore $\hat{s}_{ip}$ is not a distance metric). Clearly, such matrix can be used to derive a partition of observations through a suitable clustering technique.

Once a partition is obtained, we assume that we can identify ${k}$ pivots, $i_1,\ldots,i_{{k}}$, one for each group, which are (pairwise) separated with (posterior) probability one (that is, the posterior probability of any two of them being in the same group is zero). Ideally, the ${k}\times{k}$ submatrix of $C$ with only the rows and columns corresponding to $i_1,\ldots,i_{k}$, will be the identity matrix. 
Such units are used to identify the groups and to relabel the chains in the following way: for each $h=1,\ldots, H$ and $j=1,\ldots,k$, set
\begin{eqnarray}
\ca{\mu_j}  =  \ca{\mu_{\ca{z_{i_j}}}},\\
\ca{\eta_j}  =  \ca{\eta_{\ca{z_{i_j}}}},\\
\ca{z_i} =   j  \mbox{ for } i:\ca{z_i}=\ca{z_{i_j}}.
\label{eq:relabelZ}
\end{eqnarray}
The applicability of this strategy requires the existence of the pivots, which is not guaranteed; moreover, the  identification of pivots may be difficult, and the methods to detect them are central to the procedure (see the discussion in \cite[Section 3.1]{egidi2018relabelling} on some circumstances
in which the pivots do not exist).
Two extreme cases could occur: ($c_1$) each iteration of the chain may imply a 
number of non-empty groups strictly lower than $k$; ($c_2$) even if $k$ non-empty groups are available, however, there
may not be $k$ perfectly separated units. We then perform
the pivot relabelling discarding all those iterations that fall into one of the two categories, $c_1$ or $c_2$; the resulting restricted chain may be then seen as a sample from
an approximation of the posterior conditional to being $k$
non-empty groups. In what follows, the operational assumption is that the number of pivots should be the same as the number of filled components in the mixture model.

\subsection{Pivotal methods}
\label{subsec3}

In what follows, we assume that $k$ distinct pivotal units do exist and consider some criteria to identify them starting from the entries of the dissimilarity matrix defined above. Such criteria aim at finding those units that are as far as possible from units that might belong to other groups and/or as close as possible
to units that belong to the same group, according to a suitable objective function. Specifically, the pivot  for group $j$,  ${i^*}\in\mathcal K_j$, is chosen so that it maximizes (a) the global within similarity or (b) the difference between global within and between similarities:
\begin{align}
	\label{eq:maxmeth}
	\text{(a)} & \sum_{p\in\mathcal K_j} c_{{i^*}p},  \quad \\
	\text{(b)} & \sum_{p\in\mathcal K_j} c_{{i^*}p} - \sum_{p\not\in\mathcal K_j} c_{{i^*}p}.\nonumber
\end{align}
An alternative strategy is to find  ${i^*}\in\mathcal K_p$ that minimizes (c) the global similarity between one group and all the others
\begin{align}
	\label{eq:minmeth}
	\text{(c)} & \sum_{p\not\in\mathcal K_j} c_{{i^*}p}.
\end{align}
In the next section, we will refer to methods (a), (b) and (c) as {\tt maxsumint}, {\tt maxsumdiff} and {\tt minsumnoint}, respectively. A different method for detecting pivotal units is the Maxima Units Search (MUS), introduced in \cite{egidi2018maxima} as an algorithm for extracting identity submatrices of small rank from large and sparse matrices. The MUS algorithm is applied to the matrix  $\hat{C}=(\hat{c}_{ip})$, which is expected to contain a non-negligible number of zeros corresponding to pairs of units belonging to different groups. The procedure then seeks those units, among a pre-specified number of candidate pivots, whose corresponding rows contain more zeros compared to all other units. As a result, the MUS chooses as pivots the units--one for each group--that yield the higher number of identity submatrices of rank $k$. In practice, such matrices may contain few nonzero elements off the diagonal. It is worth noting that the MUS algorithm is in general computationally more demanding than criteria (a)--(c), since it does not rely upon a maximization/minimization step but requires an iterative investigation of the  structure of the similarity matrix $\hat{C}$. Despite the computational complexity for large $n$ and $k$, it has proved to give satisfactory results as a pivot identification technique \cite{egidi2018relabelling}.

In the rest of the paper, the pivotal methods (a), (b), (c) and MUS will be exploited in the context of Bayesian Gaussian mixture models and marginally for $k$-means robust seeding. However, the identification of pivotal units based on a co-association matrix could also be useful in other contexts, such as Dirichlet process mixtures (see Section \ref{dirichlet}) and classification rules. Some other applications of pivotal methods are left for future research.

\section{The R package pivmet} 
\label{sec:pivmet}

The {\tt pivmet} \texttt{R} package provides a simple framework to (i) fit univariate and multivariate mixture models according to a Bayesian flavor and select the pivotal units, via the {\tt piv\_MCMC} function; (ii) perform the relabelling step described in Section~\ref{sec:models} via the {\tt piv\_rel} function.

\subsection{MCMC sampling} 
\label{subsec1}

In this section, we describe how to perform MCMC sampling and pivotal units detection with the {\tt piv\_MCMC} function, whose main arguments are the following ones:

\begin{itemize}
\item {\tt{y}} $n$-dimensional  vector for univariate data or $n\times d$ matrix for multivariate data.
\item {\tt k} The number of desired mixture components.
\item {\tt nMC} The number of Markov Chain Monte Carlo (MCMC) iterations for the {\tt JAGS/Stan} execution.
\item {\tt priors} Input hyperparameters for the priors, specified as a names' list. Priors are chosen as weakly informative. For univariate mixtures, the specification is the same as the function {\tt{BMMmodel()}} of the {\tt{bayesmix}} package \cite{bayesmix} if {\tt{software="rjags"}}.  
\item {\tt{piv.criterion}} The pivotal criterion used for identifying one pivot for each group. Possible choices are: {\tt{"MUS"}}, {\tt{"maxsumint"}}, {\tt{"minsumnoint"}}, {\tt{"maxsumdiff"}}. The default method is {\tt{"maxsumdiff"}}.
\item {\tt{clustering}} The algorithm adopted for partitioning the $n$ observations into $k$ groups,the reference partition $\mathcal{P}$. Possible choices are {\tt{"diana"}} (default) or {\tt{"hclust"}} for divisive and agglomerative hierarchical clustering, respectively.
\item {\tt{software}} The selected MCMC method to fit the model: {\tt{"rjags"}} for the JAGS method, {\tt{"rstan"}} for the Stan method.
Default is {\tt{"rjags"}}.
\item {\tt{burn}} burn The burn-in period (only if method {\tt{"rjags"}} is selected).
\item {\tt{chains}} A positive integer specifying the number of Markov chains. The default is 4. 
\item  {\tt{cores}} The number of cores to use when executing the Markov chains in parallel (only if
{\tt{"rstan"}} is selected). Default is 1.
\end{itemize}

The following values are returned:

\begin{itemize}
\item \texttt{true.iter} The number of MCMC iterations for which the number of JAGS/Stan non-empty groups exactly coincides with the pre-specified number of groups \texttt{k}.
\item \texttt{groupPost} The  \texttt{true.iter}$\times n$ matrix with values from $1$ to $k$ indicating the post-processed group allocation vector.
\item \texttt{mcmc\_mean} If \texttt{y} is a vector, a \texttt{true.iter}$\times k$ matrix with the post-processed MCMC chains for the mean parameters; if \texttt{y} is a matrix, a \texttt{true.iter}$\times d\times k$ array with the post-processed MCMC chains for the mean parameters.
\item \texttt{mcmc\_sd} If \texttt{y} is a vector, a \texttt{true.iter}$\times k$ matrix with the post-processed MCMC chains for the sd parameters; if \texttt{y} is a matrix, a \texttt{true.iter}$\times d$ matrix with the post-processed MCMC chains for the sd parameters.
\item \texttt{mcmc\_weight}  A \texttt{true.iter}$\times k$ matrix with the post-processed MCMC chains for the weights parameters.
\item \texttt{grr} The vector of cluster membership returned by \texttt{"diana"} or \texttt{"hclust"}.
\item \texttt{pivots}	The vector of indices of pivotal units identified by the selected pivotal criterion.
\item \texttt{model}	 The JAGS/Stan model code. Apply the \texttt{"cat"} function for a nice visualization of the code.
\item \texttt{stanfit} An object of S4 class \texttt{stanfit} for the fitted model \\(only if \texttt{software="rstan"}).
\end{itemize}

If \texttt{software="rjags"}, the function performs JAGS sampling using the \texttt{bayesmix} package for univariate Gaussian mixtures, and the \texttt{runjags} package \cite{denwood2016} for multivariate Gaussian mixtures. If \texttt{software="rstan"}, the function performs Hamiltonian Monte Carlo (HMC) sampling using the \texttt{rstan} package.

Depending on the selected software, the model parametrization changes in terms of the prior choices. Precisely, the JAGS philosophy with the underlying Gibbs sampling is to use noninformative priors, and conjugate priors are preferred for computational speed. Conversely, Stan adopts weakly informative priors \cite{gelman2008weakly, gelman2006prior}, with no need to explicitly use the conjugacy.
For univariate mixtures, the model formulation is
\begin{equation}
y_i \sim \sum_{j=1}^k \eta_j \mathcal{N}(\mu_j, \sigma_j^2), \ \ i=1,\ldots,n.
\label{eq:fishery} 
\end{equation}
If \texttt{software="rjags"}
 the specification is the same as for the function \texttt{BMMmodel} of the \texttt{bayesmix} package:
\begin{eqnarray}
\begin{split}
\mu_j \sim & \mathcal{N}(\mu_0, B_0^2)\\
  \sigma_j \sim & \mbox{invGamma}(\nu_0/2, \nu_0S_0/2)\\
  \eta \sim & \mbox{Dirichlet}(\bm{\alpha})\\
  S_0 \sim & \mbox{Gamma}(g_0/2, g_0G_0/2),
  \end{split}
 \label{jags:prior} 
\end{eqnarray}
with default values: $B_0=10$ for the standard deviation, $\nu_0 =20$, $g_0 = 10^{-16}$, $G_0 = 10^{-16}$, and $\bm{\alpha}=(1,1,\ldots,1)$ is a vector of $k$ components. The users may specify their own hyperparameters via the \texttt{priors} argument in such a way:

\begin{Code}
   priors = list(mu_0 = 1, B0inv = 0.1, alpha =  rep(2,k)).
\end{Code}
Note that the \texttt{B0inv} is equal to $1/B_0$ from equation \eqref{jags:prior}. 
When \texttt{software="rstan"}, the prior specification is:
\begin{eqnarray}
\begin{split}
  \mu_j & \sim \mathcal{N}(\mu_0, B_0^2)\\
  \sigma_j & \sim \mbox{Lognormal}(\mu_{\sigma}, \tau_{\sigma})\\
  \eta_j & \sim \mbox{Uniform}(0,1),
  \end{split}
 \end{eqnarray}
where the vector of the weights,  $\boldsymbol{\eta}=(\eta_1,\ldots,\eta_k)$, is a $k$-simplex. Default hyperparameters values are $\mu_0=0, B_0=10, \mu_{\sigma}=0, \tau_{\sigma}=2$, which can be modified in the following way:

\begin{Code}
   priors = list(mu_phi = 0, sigma_phi = 1, B0inv = 0.1, ...).
\end{Code}
 
For multivariate mixtures, let $\bm{y}_i \in \mathbb{R}^d$ and assume that
\begin{equation}
\bm{y}_i \sim \sum_{j=1}^{k}\eta_j \mathcal{N}_d(\bm{\mu}_j, \bm{\Sigma}_j),\ \ i=1,\ldots,n,
\label{eq:bivariate}
\end{equation}
where $\bm{\mu}_j \in \mathbb{R}^d$ and $\bm{\Sigma}_j$ is a $d\times d$ positive definite covariance matrix. If \texttt{software="rjags"}  the prior specification for the parameters in \eqref{eq:bivariate} is the following:
\begin{align}
\begin{split}
\bm{\mu}_j  \sim & \mathcal{N}_2(\bm{\mu}_0, S_2)\\
 \bm{\Sigma}_j^{-1} \sim & \mbox{Wishart}(S_3, d+1)\\
\eta \sim & \mbox{Dirichlet}(\bm{\alpha}),
\end{split}
\end{align}
where  $\bm{\alpha}$ is a $k$-dimensional vector and $S_2$ and $S_3$ are positive definite matrices. By default, $\bm{\mu}_0=\bm{0}$, $\bm{\alpha}=(1,\ldots,1)$ and $S_2$ and $S_3$ are diagonal matrices, with diagonal elements 
equal to $10^5$. The user may specify alternative values for the hyperparameters
$\bm{\mu}_0, S_2, S_3$ and $\bm{\alpha}$ via \texttt{priors} argument in such a way:

\begin{Code}
   priors = list(mu_0 = c(1,1), S2 = ..., S3 = ..., alpha = ...)
\end{Code}
with the constraint for $S_2$ and $S_3$ to be positive definite, and $\bm{\alpha}$ a vector of dimension $k$ with nonnegative elements. When \texttt{software="rstan"}, the prior specification is:
\begin{align}
\begin{split}
 \bm{\mu}_j  \sim & \mathcal{N}_2(\bm{\mu}_0, LDL^{T})\\
 L \sim & \mbox{LKJ}(\epsilon)\\
D_{h} \sim & \mbox{HalfCauchy}(0, \sigma_d), \ h=1,\ldots,d.
 \end{split}
 \end{align}
The covariance matrix  for $\bm{\mu}_j$  is expressed in terms of the LDL decomposition as $LDL^{T}$,
a variant of the classical Cholesky decomposition, where $L$ is a $d \times d$
lower unit triangular matrix and $D$ is a $d \times d$ diagonal matrix.
 The Cholesky correlation factor $L$ is assigned a LKJ prior with $\epsilon$ degrees of freedom,  which, combined with priors on the standard deviations of each component, induces a prior on the covariance matrix; as $\epsilon \rightarrow \infty$ the magnitude of correlations between components decreases, whereas $\epsilon=1$ leads to a uniform prior distribution for $L$.  By default, the hyperparameters are $\bm{\mu}_0=\bm{0}$, $\sigma_d=2.5, \epsilon=1$. The user may propose some different values with the argument:
\begin{Code}
   priors = list(mu_0=c(1,2), sigma_d = 4, epsilon = 2)
\end{Code}

Clearly, different samplers can yield different solutions in terms of posterior estimates and convergence diagnostics. We think that this is perfectly 
acceptable and we believe that the choice between the Gibbs sampling performed by JAGS software and the HMC returned by the Stan ecosystem should be driven 
by the users' preferences in terms of their individual experiences. Gibbs sampler is usually faster and requires less tuning, however HMC fit provides more diagnostics measures and is better suited for capturing the geometry of the posterior 
distribution in big dimensions. In our opinion, JAGS performance is overall good for univariate and bivariate mixtures, whereas Stan can be preferable when $d>2$. Moreover, the choice of the priors is completely different in the two approaches: in this package, default hyperparameters have been chosen upon some sensitivity checks and in compliance with some guidelines provided by vignettes and manuals of both JAGS and Stan.

Finally, it is worth mentioning that the value \texttt{true.iter} returned by \texttt{piv\_MCMC} is the length of the MCMC chains after applying criterion ($c_1$) mentioned in Section \ref{sec:relabelling}, and represents the number of
MCMC iterations for which the number of JAGS/Stan groups exactly coincides with $k$: for such a reason, this number is less than or equal than the input argument \texttt{nMC}. The value \texttt{final\_it} provided by the \texttt{piv\_rel} function in the next Section,  yields the final number of valid MCMC iterations, and is obtained after performing ($c_2$) criterion on \texttt{true.iter}.

\subsection{Pivotal methods and relabelling}
\label{subsec2}

After MCMC sampling, a clustering algorithm specified via the argument \texttt{clustering} is applied to the units in order to find $k$ groups from the sample, i.e. the reference partition $\mathcal{P}$ defined in \eqref{eq:partition}. Then, by default, the internal function \texttt{piv\_sel} is used to obtain the pivots from criteria \texttt{"maxsumint"}, \texttt{"minsumnoint"} and \texttt{"maxsumdiff"}  of \eqref{eq:maxmeth}-\eqref{eq:minmeth}. Alternatively, when $k <5$, one can use \texttt{piv.criterion="MUS"}, which performs a sequential search of identity submatrices within the matrix $C$, and returns the pivots via the internal function \texttt{MUS}.\\

The \texttt{piv\_rel} function requires the output from \texttt{piv\_MCMC} and consists of the following argument:
\begin{itemize}
\item \texttt{mcmc} The output of the MCMC sampling from \texttt{piv\_MCMC}.
\end{itemize} 

The following values are returned:

\begin{itemize}
\item \texttt{final\_it} The final number of valid MCMC iterations
\item \texttt{rel\_mean}	The relabelled chains of the means: a \texttt{final\_it}$ \times k$ matrix for univariate data, or a \texttt{final\_it}$ \times d \times k$ array for multivariate data.
\item \texttt{rel\_sd}	The relabelled chains of the sd's: a \texttt{final\_it}$ \times k$ matrix for univariate data, or a \texttt{final\_it}$ \times d$ matrix for multivariate data.
\item \texttt{rel\_weight} The relabelled chains of the weights: a \texttt{final\_it}$ \times k$ matrix.
\end{itemize}
A graphical visualization of the new posterior estimates may be obtained via the function \texttt{piv\_plot}, which takes as input the data, the MCMC output from \texttt{piv\_MCMC} and the relabelled posterior chains output from \texttt{piv\_rel}.
 
The next section illustrates how the relabelling algorithm works, starting from the simulation of an artificial dataset via the \texttt{piv\_sim} function in the \texttt{pivmet} package. In particular, \texttt{piv\_sim} allows to simulate  a sample of size $n$ from a mixture of Gaussian distributions with suitable parameters, with $k$ components.

\subsection{Mixtures of bivariate Gaussian distributions}

For simulating bivariate Gaussian data, the function \texttt{piv\_sim} assumes that, conditional on being in group $j$ ($j=1, \dots, k$), each observation is drawn from one of two possible Gaussian distributions (sub-groups), using a vector of weights $\bm{w}=(w_1, w_2)$ specified by the argument \texttt{W}. This allows for a broader within-groups heterogeneity, similarly to what happens for the `spike-and-slab' approach \cite{spike}, where one sub-component is flat around $\mu_j$ (the slab part), and the other one is concentrated and peaked (the spike part). The \texttt{Sigma.p1} and \texttt{Sigma.p2} represent the covariance matrix for the first and second sub-group, respectively, while the argument \texttt{Mu} is the $k \times 2$ matrix of input means. To generate data from model \eqref{eq:bivariate}, we can fix either $w_1=0, w_2=1$ or $w_1=1, w_2=0$. As an illustration, we simulate a sample of bivariate data ($n=150$) from $k=4$ groups with the following commands:

\begin{Code}
> library(pivmet)	
> library(rstan)
> set.seed(10)
> N <- 150          # sample size
> k <- 4            # number of mixture components
> D <- 2            # data dimension
> nMC <- 5000       # MCMC iterations
> M1  <- c(-.5,8)
> M2  <- c(25.5,.1)
> M3  <- c(49.5,8)
> M4  <- c(25,25)
> Mu  <- rbind(M1,M2,M3,M4)  # Input mean
> Sigma.p1 <- diag(D)        # Cov. matrix first subgroup
> Sigma.p2 <- (14^2)*diag(D) # Cov. matrix second subgroup
> W <- c(0.2,0.8)            # Weights
> sim <- piv_sim(N = N, k = k , Mu = Mu, Sigma.p1 = Sigma.p1, 
+	Sigma.p2 = Sigma.p2, W = W)  # simulated data
\end{Code} 
Then, the \texttt{piv\_MCMC} and \texttt{piv\_rel} functions, with the argument \texttt{software="rstan"} for the first one, are used. 

\begin{figure}[t!]
\centering
\includegraphics[width=.8\linewidth]{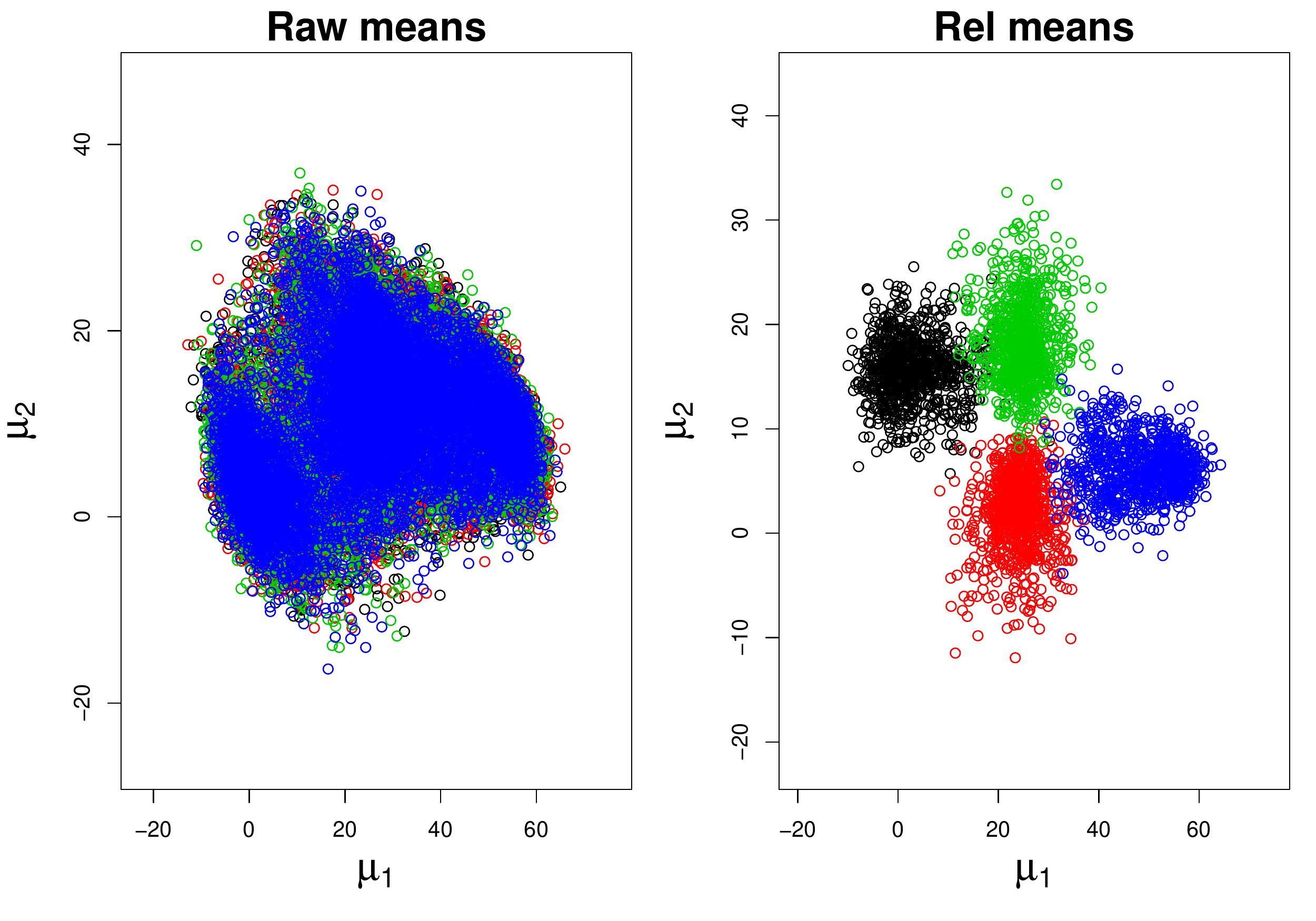}
\caption{Bivariate mixture data: traceplot for the mean parameters obtained via the \texttt{rstan} option, 5000 HMC iterations.}
\label{bivariate_chains}
\end{figure}

\begin{figure}[t!]
\centering
\includegraphics[width=.8\linewidth]{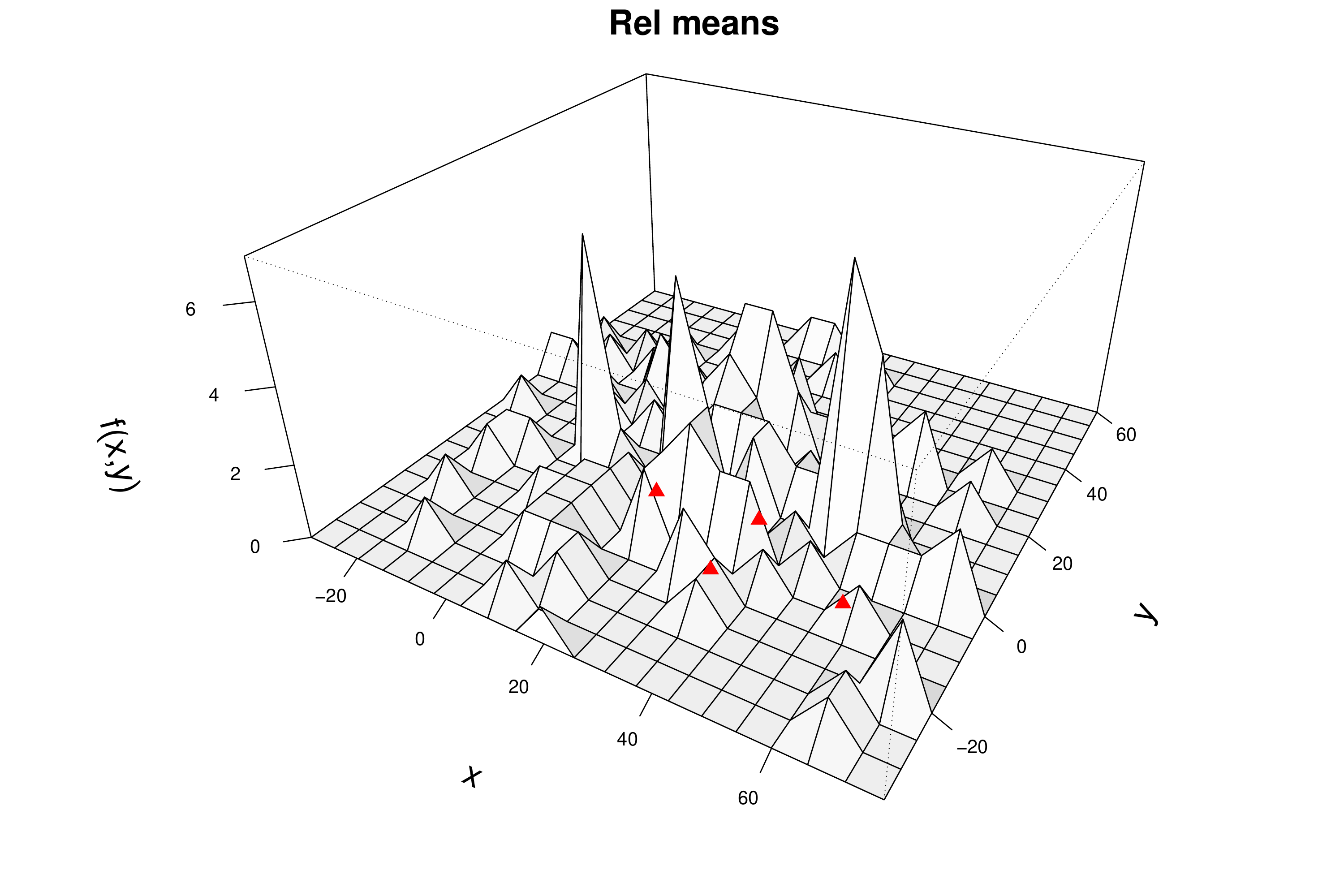}
\caption{Bivariate mixture data: 3D histogram and posterior estimates from the relabelled chains (red triangles).}
\label{bivariate_hist}
\end{figure}

\begin{Code}
> res <- piv_MCMC(y = sim$y, k = k, nMC =nMC, piv.criterion="MUS", 
+	software="rstan")  # HMC sampling
> rel <- piv_rel(mcmc = res)  # relabelling
\end{Code}
The \texttt{piv\_plot} function with the argument \texttt{par = "mean"} yields the bivariate traceplot chains for each mean component $\mu_{j,1}$ and $\mu_{j,2}$ (Figure~\ref{bivariate_chains}); the same function with the argument \texttt{type = "hist"} produces a 3d histogram of the simulated data along with the relabelled posterior estimates for each $\bm{\mu}_j$, marked with triangle red points (Figure~\ref{bivariate_hist}).

\begin{Code}
> piv_plot(y=sim$y, mcmc=res, rel_est = rel, par ="mean", 
			type="chains")  # traceplots
> piv_plot(y=sim$y, mcmc=res, rel_est = rel, type="hist")
\end{Code}

Clearly, label switching has occurred and the relabelling algorithm fixed it, by isolating the four bivariate high-density regions.
		
\subsection{Consensus clustering based on pivots} 
\label{subsec2}
Ensembles methods have recently emerged as a valid alternative to conventional clustering techniques \cite{JMLR02}, since they allow to summarize the information coming from multiple
clusterings. These can be obtained, for instance, by applying different clustering algorithms, by using the same algorithm with different parameters or initializations, or by adopting  different dissimilarity measures. The resulting clustering ensemble is then used to obtain a \emph{consensus} partition, according to the idea of evidence accumulation, i.e., by viewing each clustering result as an independent evidence of data structure.


Given  a set of $n$ observations  $(y_{1},\ldots,y_{n})$, a common approach uses the ensemble to derive a new pairwise similarity matrix, or co-association matrix, by taking the co-occurrences of pairs of points in the same group across all partitions, i.e. $c_{ij}=n_{ij}/H$, where $n_{ij}$ is the number of times the pair $(y_i,y_j)$ is assigned to the same cluster among the $H$ partitions \cite{fredJain05}.

In \cite{kmeans}, the idea of using pivots for performing data clustering is presented, and a modified version of the well-known $k$-means algorithm  is proposed. In particular, the starting point is the co-association matrix obtained from multiple runs of the $k$-means algorithm with initial random seeds. Such matrix is given as input for the MUS procedure, and the resulting pivots are regarded to as cluster centers for the $k$-means run yielding the consensus partition. Such approach can be viewed as a strategy for careful seeding which may improve the validity of the final configuration, and overcome well-know limitations of the standard method; for a different approach to careful seeding see \cite{kmeans07}. The modified $k$-means clustering which uses a pivot-based initialization
step is implemented via the function \texttt{piv\_KMeans}, where the approach described in  \cite{kmeans} is extended in order to allow the user to choose the pivotal criterion applied to the co-association matrix. 
The general usage is:

\begin{Code}
piv_KMeans(x, centers, alg.type = "KMeans", method = "average",
 	     piv.criterion = c("MUS", "maxsumint", 
 	    "minsumnoint", "maxsumdiff"),
              H = 1000, iter.max = 10, 
              num.seeds = 10, prec_par = 10)
\end{Code} 
By default \texttt{piv\_KMeans} function takes the data, in either matrix or data frame format, and obtains a partition $\mathcal{P}$ of \texttt{x} into a user-specified number of groups (\texttt{centers}) via the $k$-means algorithm. The selected pivotal identification method requires such partition as input, in order to find $k$ (distinct) pivots from the co-association matrix resulting from  $H=1000$ runs of $k$-means with different starting random seeds. By default, if \texttt{centers < 5}, MUS algorithm is used;  otherwise, the default pivotal method is \texttt{maxsumint} (see Section~\ref{sec:relabelling}).
Finally, the function returns the clustering obtained by using the pivots as initial group centers in function \texttt{kmeans()} of  the \texttt{stats} package. By setting \texttt{alg.type="hclust"}, agglomerative hierarchical clustering  is used with usual agglomeration methods for obtaining the reference partition $\mathcal{P}$. In such case, method \texttt{"average"} is adopted by default; other possible choices for the \texttt{method} argument are the ones implemented in the \texttt{hclust()} {R} function. 
The maximum number of iterations and the number of different starting random seeds, which only apply when \texttt{alg.type = "KMeans"}, have default value 10, while the number of candidate pivots to be considered for \texttt{MUS} procedure is specified by the argument \texttt{prec\_par}.
Note that the sensitivity of the proposed method to the choice of the method used to obtain the reference partition is part of ongoing research, however preliminary investigation shows that hierarchical clustering provides overall good results in terms of pivots separation for arbitrary cluster shapes. 

\begin{figure}[ht]
	\begin{center}
	\includegraphics[width=.6\linewidth]{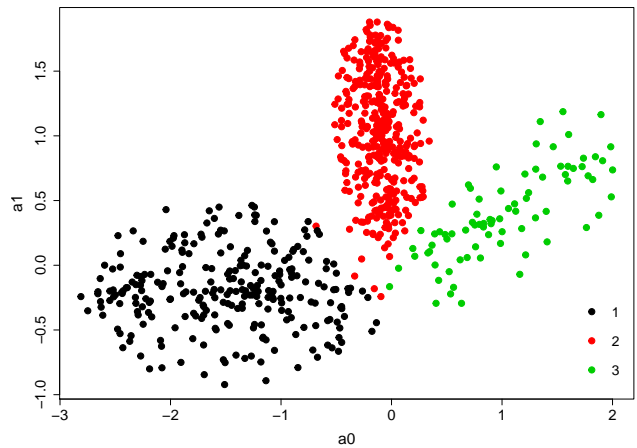}
	\caption{Clustering of \texttt{2d-3c-no123} dataset obtained via the \texttt{pivKMeans} function with $k=3$ and default pivotal method \texttt{MUS}.}
    \label{pivkm}       
 \end{center}
\end{figure}

As an illustration, a cluster analysis of artificial data available from Tomas Barton's clustering benchmark is shown below (data available at \url{https://github.com/deric/clustering-benchmark}). We import the \texttt{2d-3c-no123} dataset with function \texttt{read.arff} of the \texttt{foreign} package \cite{foreign}, which consists of three groups of bivariate data (with sizes 264, 370, 81, respectively). \texttt{piv\_KMeans} is applied to the two first columns, which contains the data points, with option \texttt{alg.type="hclust"}:

\begin{Code}
> library(foreign)
> library(mclust)
> file.n <-  "https://raw.githubusercontent.com/deric/clustering-
	benchmark/master/src/main/resources/datasets/artificial/
	2d-3c-no123.arff"
> data <- read.arff(file.n)
> x <- data[, 1:2]
> pkm <- piv_KMeans(x, 3, alg.type = "hclust")
\end{Code}
The output can be inspected by calling the object \texttt{pkm} (not shown here). Clustering results can be visualized via \texttt{plot()} (see Figure~\ref{pivkm}). We evaluate the resulting partition by comparison with the true cluster labels, and we compute the Adjusted Rand Index (ARI) implemented in \texttt{mclust} \cite{mclust}, as an external validity measure of clustering  quality. 

\begin{Code}
> plot(x, col = pkm$cluster, pch=19)
> legend("bottomright", legend=c("1", "2", "3"), 
	pch=19, col=c(1:3))
> table(pkm$cluster, as.numeric(data[, 3]))
## 
##      1   2   3
## 1  257   0   0
## 2    6 370   2
## 3    1   0  79
> adjustedRandIndex(pkm$cluster, as.numeric(data[, 3]))
## 
## [1] 0.959636
\end{Code}
		
\subsection{Dirichlet Process Mixtures}
\label{dirichlet}

In some situations, the choice of an appropriate prior distribution for the group means $\mu_j$ as in \eqref{jags:prior} is a troublesome issue, particularly when the number of observations is small. In this case,
adopting a nonparametric Dirichlet Process Mixture Model (DPMM) \cite{escobar1995bayesian, neal2000markov} specification for the prior on $\mu_j$ avoids an inappropriate parametric form. Assuming that the parameter vector is given by $\theta=(\mu, \sigma)$, DPMM has the following form:

\begin{align}
\begin{split}
y_i \sim &\  K(y_i|\theta_i),\ i=1,\ldots,n \\
\theta_i \sim &\ F\\
F \sim &\ \text{DP}(\alpha, G),
\end{split}
\end{align}
where $K(\cdot)$ is a parametric kernel function which is usually continuous, $F$ is an unknown probability distribution, $\text{DP}$ is the nonparametric Dirichlet process prior with concentration parameter $\alpha$ and \emph{base measure} $G$, which encapsulates any prior knowledge about $F$ \cite{ferguson1973bayesian}. A common choice for $K(\cdot)$ is a Gaussian mixture model, so that $K(y_i|\theta_i)= \mathcal{N}(\mu_i, \sigma^2_i)$.
The DPMM sorts
the data into clusters, corresponding to the mixture components. Thus, it may be seen as an infinite dimensional mixture model which generalizes finite mixture models. Thus, pivotal units detection may be quite relevant for this class of models in order to identify distinct groups characteristics.

\begin{figure}
\centering
\includegraphics[scale=0.4]{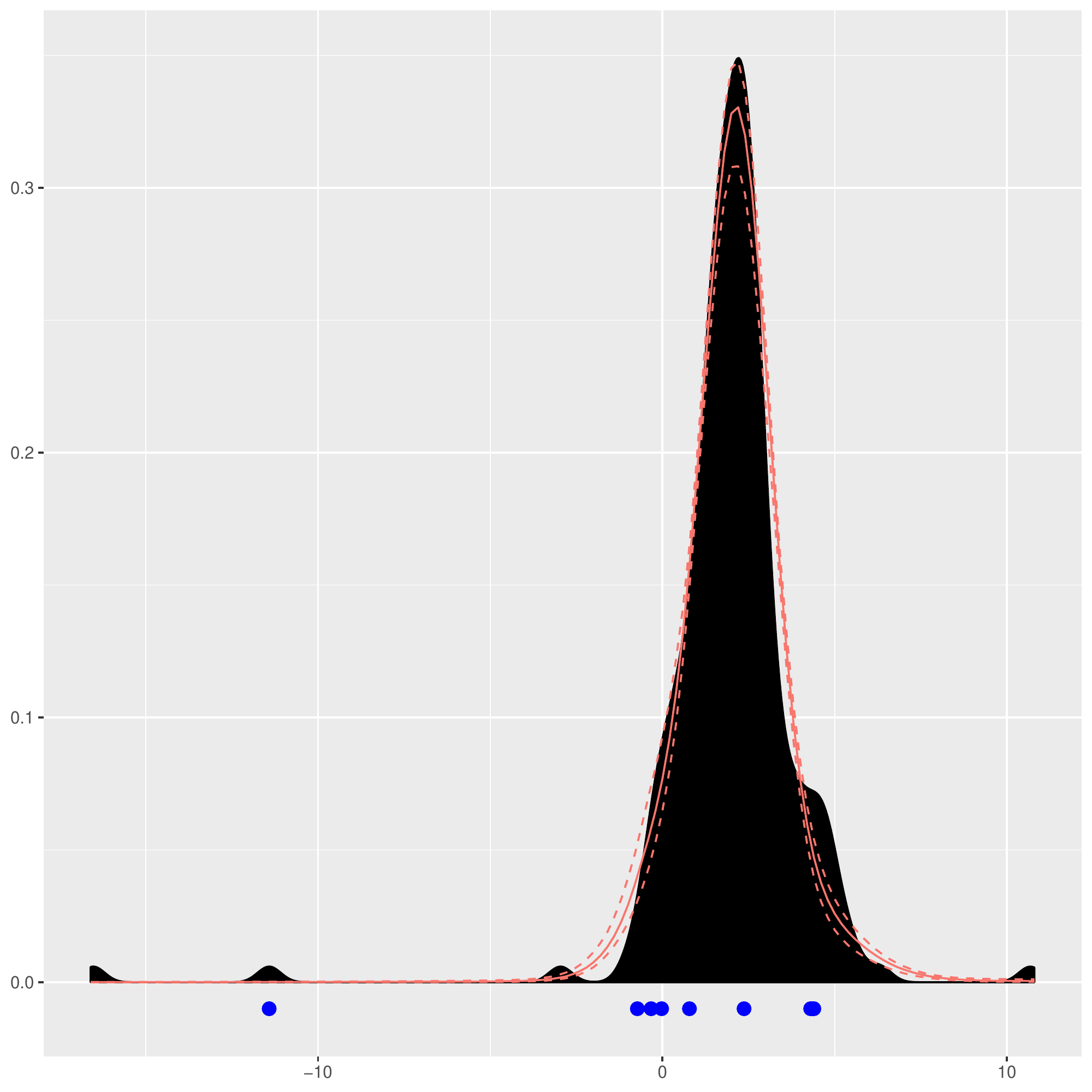}
\caption{Posterior density estimation (red line) for a sample of $n=200$ data points from a student$-t$ distribution (\texttt{dirichletprocess} package). Blue points below the $x$-axis denote the pivotal units provided by \texttt{maxsumdiff} criterion of the \texttt{piv\_sel} function.}
\label{dppm}
\end{figure}

For illustration purposes only, we use the \texttt{dirichletprocess} package \cite{ross2018dirichletprocess} to generate a simulated example. We generate $n=200$ data from a student$-t$ distribution with 3 degrees of freedom and we use the \texttt{Fit} function from the same package to draw posterior samples for $\mu_1,\mu_2,\ldots,\mu_k$ via the Chinese Restaurant Process sampler \cite{neal2000markov}. In DPMM framework, the number of clusters $k$ is unknown, and is estimated from the posterior distribution:

\begin{Code}
> library(dirichletprocess)
> library(ggplot2)
> set.seed(1234)
> n <- 200 # sample size
> nMC <- 1000 # MCMC iterations
> y <- rt(n, 3) + 2 #generate sample data
> dp <- DirichletProcessGaussian(y) 
> dp <- Fit(dp, nMC) # MCMC sampling (dirichletprocess)
> dp$numberClusters # number of "non-empty" clusters
## 9
\end{Code}
The number of clusters from the posterior distribution is $k=9$.
Now, we can build the estimated co-association matrix $\hat{C}$, whose entries are defined in \eqref{eq:Cmatrix} across the $H=1000$ MCMC iterations:

\begin{Code}
> C_array <- array(1, dim = c(nMC, n, n)) # co-association array
> for (h in 1:nMC){
    for (i in 1:(n-1)){
       for (j in (i+1):n){
      	if (dp$labelsChain[[h]][i]==dp$labelsChain[[h]][j]){
      		C_array[h,i,j] <- 1
      		}else{
      		C_array[h,i,j] <- 0  
      }
    }
  }
}
> C <- apply(C_array, c(2,3), mean) # co-association matrix
\end{Code}
We are ready to extract the pivots from $\hat{C}$ by using the \texttt{piv\_sel} function, according to the methods (a), (b), (c) in \eqref{eq:maxmeth} and \eqref{eq:minmeth}. We use the reference partition $\mathcal{P}$ provided by the DPMM fit:

\begin{Code}
> piv_selection <- piv_sel(C = C, clusters = dp$clusterLabels) 
			  # pivotal methods
> piv_index <- piv_selection$pivots[,3] # maxsumdiff
> df_piv <- data.frame(x=y[piv_index], 
                     y = rep(-0.01, dp$numberClusters))
> plot(dp)+geom_point(aes(x = x, y = y), data = df_piv,
                    colour = "blue", size = 1.5) # ggplot2
\end{Code}
Figure \ref{dppm} represents posterior density estimation for the simulated dataset along with the nine pivotal units (blue points) detected by the \texttt{maxsumdiff} method of the \texttt{piv\_sel} function.

\color{black}

\section{Examples}
 \label{sec:illustrations}
 
\subsection{Fishery Data}
\begin{figure}[t!]
\centering
\includegraphics[width=.8\linewidth]{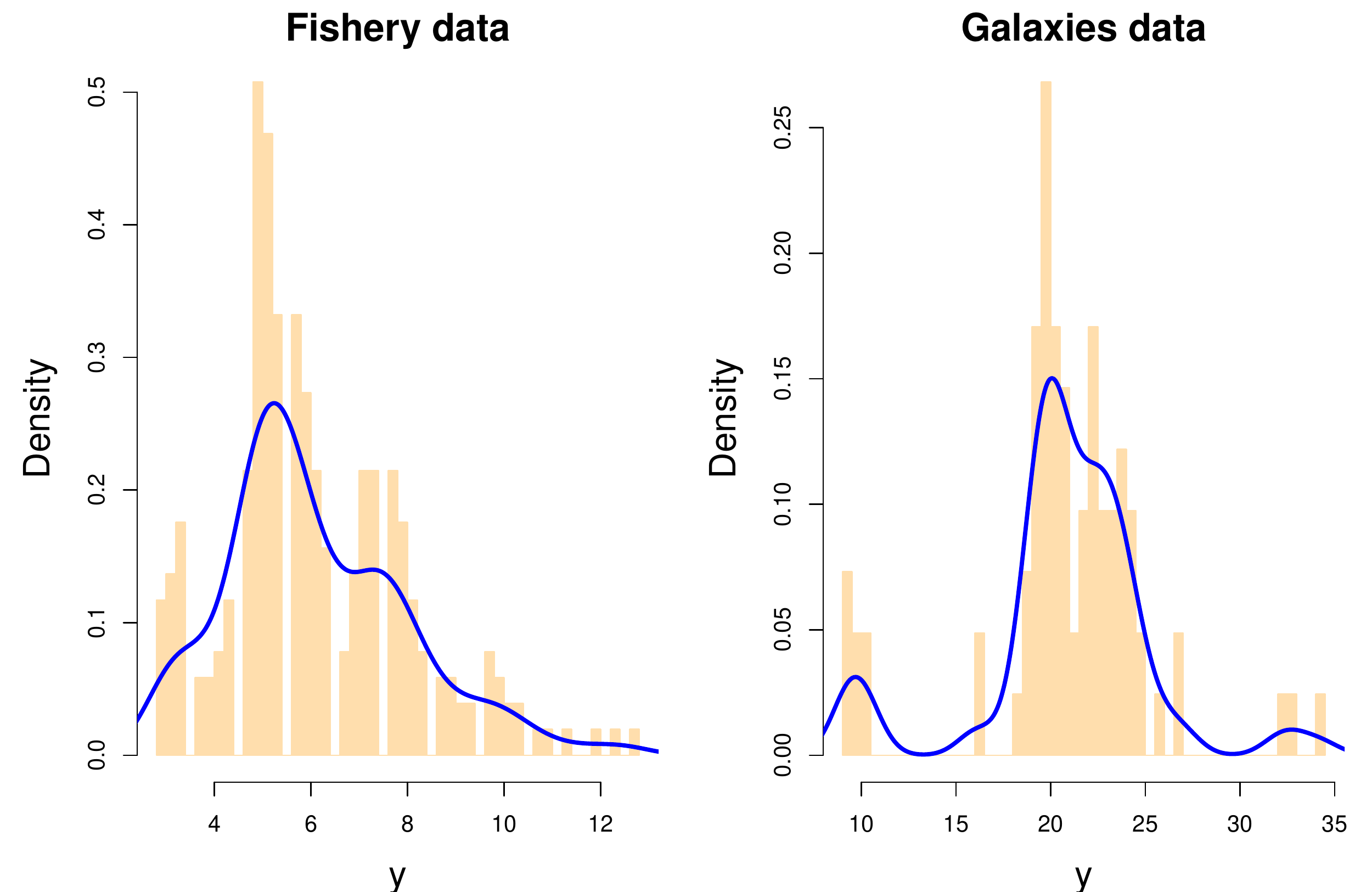}
\caption{Histograms of the Fishery and Galaxies data. The blue line represents the estimated kernel density.}
\label{hist}
\end{figure}

The Fishery dataset in the \texttt{bayesmix} package has been previously used by \cite{titterington1985statistical} and \cite{papastamoulis2016label}. It consists of 256 snapper length measurements (see left plot of Figure~\ref{hist} for the data histogram, along with an estimated kernel density). Analogously to some previous works, we assume the mixture model \eqref{eq:fishery},  with $k=5$ groups, where $\mu_j$, $\sigma_j$ and $\eta_j$ are the mean, the standard deviation and the weight of group $j$, respectively. We fit our model by simulating $H=15000$ samples from the posterior distribution of $(\bm{z}, \bm{\mu}, \bm{\sigma}, \bm{\eta})$, by selecting the default argument \texttt{software="rjags"}; for univariate mixtures, the MCMC Gibbs sampling is returned by the function \texttt{JAGSrun} in the package \texttt{bayesmix}. By default, the burn-in period is set equal to half of the total number of MCMC iterations. Partial output is shown below: 
 
\begin{Code}
> library(bayesmix)
> data(fish)
> y <- fish[,1]
> k <- 5
> nMC <- 15000
> res <- piv_MCMC(y = y, k = k, nMC = nMC, burn = 0.5*nMC,
+	software = "rjags")
## 
## Call:
## JAGSrun(y = y, model = mod.mist.univ, control = control)
## 
## Markov Chain Monte Carlo (MCMC) output:
## Start = 7501 
## End = 22500 
## Thinning interval = 1 
## 
## Empirical mean, standard deviation and 95% CI for eta 
##         Mean      SD     2.5%  97.5%
## eta[1] 0.2030 0.20230 0.009156 0.5672
## eta[2] 0.2006 0.13765 0.009236 0.5198
## eta[3] 0.1674 0.12353 0.018239 0.5051
## eta[4] 0.1217 0.05231 0.071350 0.1988
## eta[5] 0.3073 0.22254 0.009916 0.5752
##
## Empirical mean, standard deviation and 95% CI for mu 
##       Mean     SD  2.5%  97.5%
## mu[1] 8.386 2.7806 3.377 12.341
## mu[2] 8.413 2.1134 5.178 12.347
## mu[3] 8.369 1.5851 5.201 10.862
## mu[4] 3.472 0.6172 3.126  5.368
## mu[5] 7.358 2.7863 5.069 12.292
##
## Empirical mean, standard deviation and 95% CI for sigma2 
##            Mean     SD   2.5%  97.5%
## sigma2[1] 0.4473 0.2510 0.1932 1.1808
## sigma2[2] 0.4387 0.2108 0.2009 1.0234
## sigma2[3] 0.4526 0.2415 0.2021 1.1223
## sigma2[4] 0.2606 0.1128 0.1280 0.5479
## sigma2[5] 0.4132 0.2442 0.2035 1.1182
\end{Code}
Firstly, the object \texttt{res\$true.iter} yields the amount of iterations in which the number of groups returned by the MCMC sampling coincides with $k$.  For instance, in this case \texttt{res\$true.iter} was equal to 7421, meaning that approximately only 1\% of the chains'  iterations has been discarded (note that there is a burn-in period of 7500 iterations). 

From the printed output of posterior estimates of $\mu_j$, it seems clear that label switching has occurred, in fact all the means are quite close to each other, with the exception of $\mu_4$. The function \texttt{piv\_rel} allows to relabel the chains and to obtain useful inferences. The function \texttt{piv\_plot} displays some graphical tools, such as traceplots (argument \texttt{type="chains"}) and histograms also showing the final relabelled means (argument \texttt{type="hist"}) for the model parameters (not shown).

\begin{figure}[t!]
\centering
\includegraphics[width=.8\linewidth]{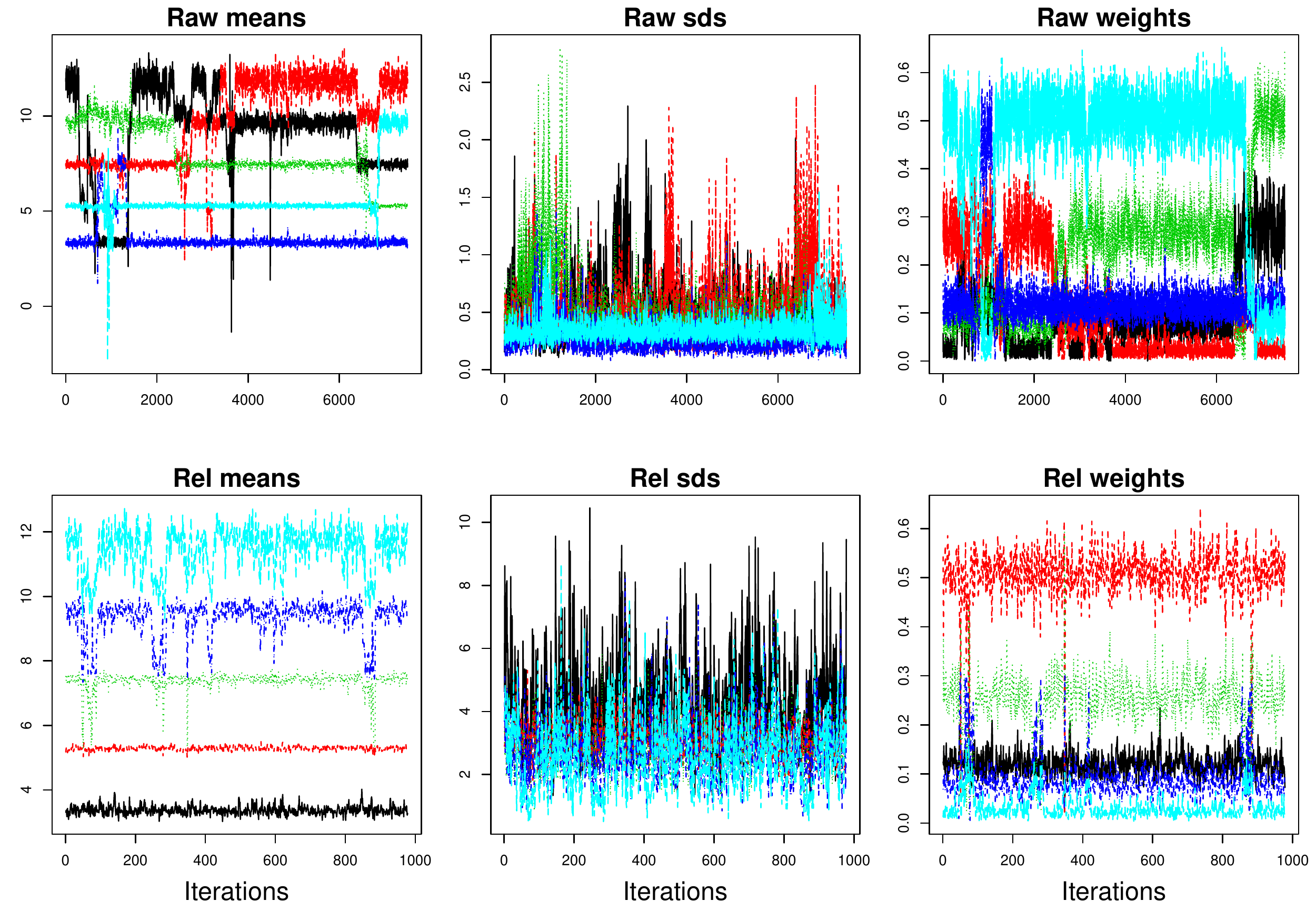}
\caption{Fishery dataset: traceplots of the parameters $(\bm{\mu}, \bm{\sigma}, \bm{\eta})$ obtained via the \texttt{rjags} option (Gibbs sampling, 15000 MCMC iterations). Top row: Raw MCMC outputs. Bottom row: Relabelled MCMC samples.}
\label{fish_chains}
\end{figure}

\begin{Code}
> rel <- piv_rel(mcmc=res)
> rel$final_it  # final number of valid MCMC iterations
## 978
> piv_plot(y = y, mcmc = res, rel_est = rel, type="chains")
## Description: traceplots of the raw MCMC chains and the 
## relabelled chains for all the model parameters:
## means, sds and weights.
## Each colored chain corresponds to one of the k 
## distinct parameters of the mixture model. 
## Overlapping chains may reveal that the MCMC sampler 
## is not able to distinguish between the components.
\end{Code}
Figure~\ref{fish_chains} displays the traceplots for the parameters $(\bm{\mu}, \bm{\sigma}, \bm{\eta})$. From the first row showing the raw outputs as given by the Gibbs sampling, we note that label switching clearly occurred. Our algorithm seems able to reorder the mean $\mu_j$ and the weights $\eta_j$, for $j=1,\ldots,k$. Of course, a MCMC sampler which does not switch the labels would be ideal, but nearly impossible to program. However, we could assess performance from different samplers by repeating the analysis above with \texttt{software="rstan"} in the \texttt{piv\_MCMC} function. We may also extract the Stan code as follows:

\begin{Code}
> library(rstan)
> res_stan <- piv_MCMC(y = y, k = k, nMC = nMC/3,
+	burn = 0.5*nMC/3, software ="rstan")
> cat(res_stan$model)

        data {
          int<lower=1> k;          // number of mixture components
          int<lower=1> N;          // number of data points
          real y[N];               // observations
          real mu_0;               // mean hyperparameter
          real<lower=0> B0inv;     // mean hyperprecision
          real mu_sigma;           // sigma hypermean
          real<lower=0> tau_sigma; // sigma hyper sd
          }
        parameters {
          simplex[k] eta;           // mixing proportions
          ordered[k] mu;            // locations of mixture components
          vector<lower=0>[k] sigma; // scales of mixture components
          }
        transformed parameters{
          vector[k] log_eta = log(eta);  // cache log calculation
          vector[k] pz[N];
          simplex[k] exp_pz[N];
              for (n in 1:N){
                  pz[n] =   normal_lpdf(y[n]|mu, sigma)+
                            log_eta-
                            log_sum_exp(normal_lpdf(y[n]|mu, sigma)+
                            log_eta);
                  exp_pz[n] = exp(pz[n]);
                            }
          }
        model {
          sigma ~ lognormal(mu_sigma, tau_sigma);
          mu ~ normal(mu_0, 1/B0inv);
            for (n in 1:N) {
              vector[k] lps = log_eta;
                for (j in 1:k){
                    lps[j] += normal_lpdf(y[n] | mu[j], sigma[j]);
                    target+=pz[n,j];
                    }
              target += log_sum_exp(lps);
                  }
          }
        generated quantities{
         int<lower=1, upper=k> z[N];
          for (n in 1:N){
              z[n] = categorical_rng(exp_pz[n]);
            }
        }
\end{Code}

\begin{figure}[t!]
\centering
\includegraphics[width=.8\linewidth]{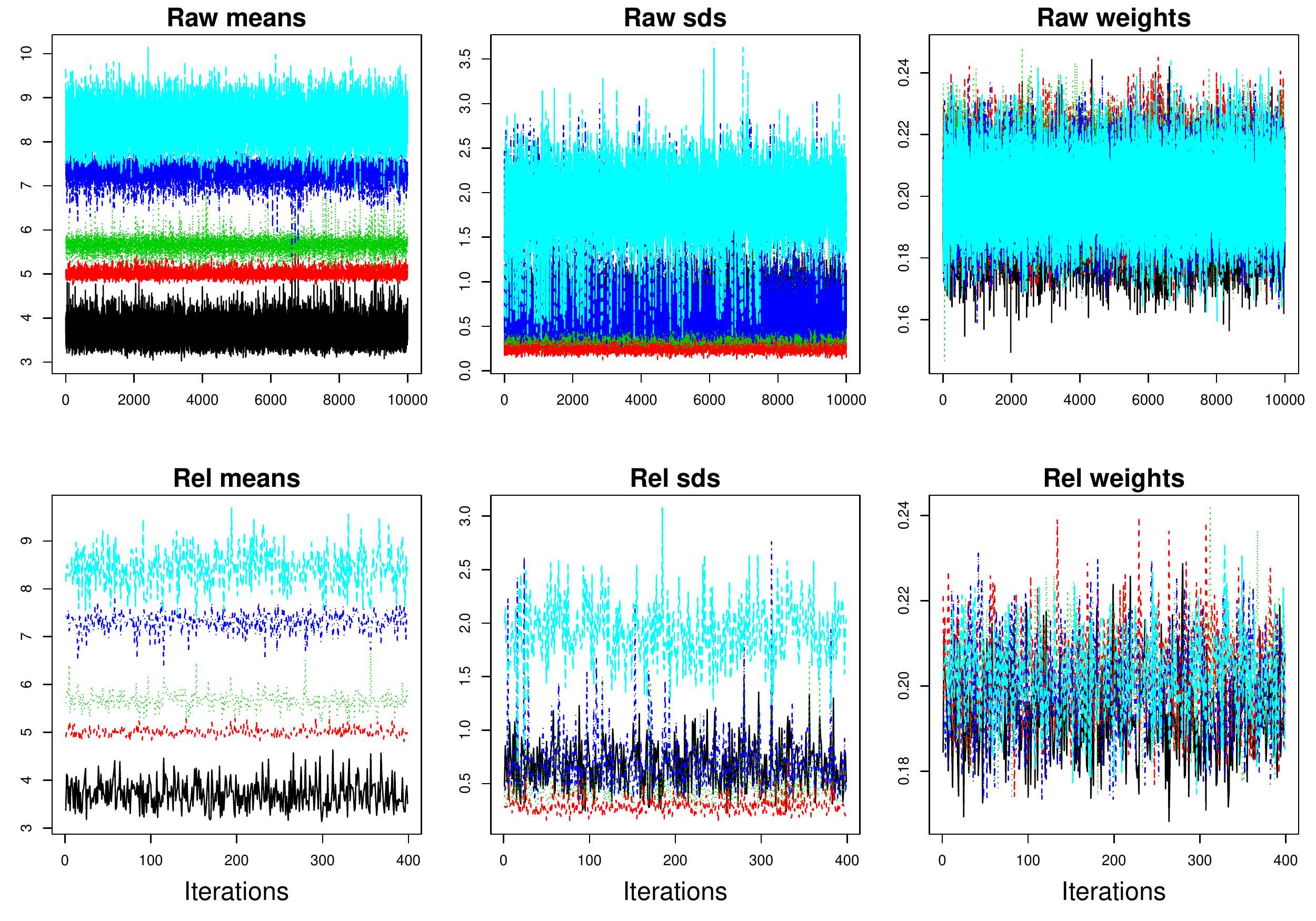}
\caption{Fishery dataset: traceplots of the parameters $(\bm{\mu}, \bm{\sigma}, \bm{\eta})$ obtained via the \texttt{rstan} option ($k=5$, $5000$  HMC iterations). Top row: Raw MCMC outputs. Bottom row: Relabelled MCMC samples.}
\label{fish_chains_stan}
\end{figure}

We can print the \texttt{stanfit} model to have a glimpse about posterior estimates and model diagnostics:

\begin{Code}
> print(res_stan$stanfit, pars = c("mu", "sigma", "eta"))
## Inference for Stan model: 0d12844c9c18f5cbe956e053b3115c3d.
## 4 chains, each with iter=5000; warmup=2500; thin=1; 
## post-warmup draws per chain=2500, total post-warmup draws=10000.
##
##         mean  se_mean   sd 2.5%  25%  50%  75% 97.5% n_eff Rhat
## mu[1]    3.70    0.01 0.29 3.23 3.48 3.67 3.88  4.33  1040 1.00
## mu[2]    5.00    0.00 0.09 4.85 4.94 4.99 5.04  5.21  2839 1.00
## mu[3]    5.66    0.00 0.17 5.27 5.57 5.67 5.76  5.93  1297 1.00
## mu[4]    7.33    0.00 0.19 6.87 7.24 7.36 7.46  7.64  3923 1.00
## mu[5]    8.45    0.01 0.43 7.59 8.15 8.45 8.75  9.30  1421 1.00
## sigma[1] 0.67    0.01 0.21 0.30 0.52 0.66 0.80  1.12  1518 1.00
## sigma[2] 0.30    0.01 0.13 0.18 0.23 0.27 0.33  0.56   358 1.01
## sigma[3] 0.46    0.00 0.16 0.29 0.37 0.43 0.51  0.81  1331 1.01
## sigma[4] 0.69    0.02 0.33 0.38 0.51 0.62 0.76  1.87   254 1.01
## sigma[5] 1.86    0.01 0.33 1.21 1.68 1.86 2.06  2.49   542 1.01
## eta[1]   0.19    0.00 0.01 0.17 0.19 0.19 0.20  0.22  5631 1.00
## eta[2]   0.20    0.00 0.01 0.18 0.19 0.20 0.21  0.23  8572 1.00
## eta[3]   0.20    0.00 0.01 0.18 0.20 0.20 0.21  0.23  8429 1.00
## eta[4]   0.20    0.00 0.01 0.18 0.19 0.20 0.21  0.22  8186 1.00
## eta[5]   0.20    0.00 0.01 0.18 0.19 0.20 0.21  0.22  8875 1.00
##
## Samples were drawn using NUTS(diag_e) at Tue Jun 09 17:42:57 2020.
## For each parameter, n_eff is a crude measure of effective sample 
## size, and Rhat is the potential scale reduction factor on split 
## chains (at convergence, Rhat=1).
\end{Code}
The chains converged (\texttt{Rhat}$<1.1$ for each parameter) and the effective sample size (\texttt{n\_eff}) does not indicate problems in the samples' autocorrelation.
The graphical results are shown in Figure~\ref{fish_chains_stan}. As may be noted from the first plot in the top row, Hamiltonian Monte Carlo (HMC) behind Stan's workflow seems definitely more suited to explore the five high-density regions without switching the group labels. However, group probabilities (third plot) and group standard deviations (second plot) overlap each other, suggesting that the perfect MCMC sampler does not exist. 

\subsection{Galaxies Data}
 
We further illustrate the usefulness of \texttt{pivmet} by considering the \texttt{galaxies} dataset in the \texttt{MASS} package \cite{mass}, consisting of  a vector of velocities in km/sec of 82 galaxies from six well-separated conic sections. Such data were firstly analyzed by \cite{roeder1990density} and then used, among the others, by \cite{richardson1997bayesian} and \cite{jasra2005markov}. The histogram of the data is shown in the right panel of Figure~\ref{hist}. Again, we assume model~\eqref{eq:fishery},
choosing $k=3$ components (as also done in \cite{stephens2000dealing}).
An informative prior $\mathcal{N}(\mu_0, 0.1^2)$ for $\mu_j$ may be appropriate, to distinguish between the group means, and specified via the \texttt{priors} argument of the \texttt{piv\_MCMC} function. To this aim, run the code:
\begin{Code}
> library(MASS)
> data(galaxies)
> y <- galaxies
> y <- y/1000
> k <- 3
> nMC <- 15000
> res <- piv_MCMC(y = y, k = k, nMC = nMC, priors=list(B0inv = 10), 
+	burn = 0.5*nMC)
> rel <- piv_rel(mcmc=res)
> piv_plot(y=y, res, rel, type="chains")
\end{Code}
%
\begin{figure}[t!]
\centering
\includegraphics[width=.8\linewidth]{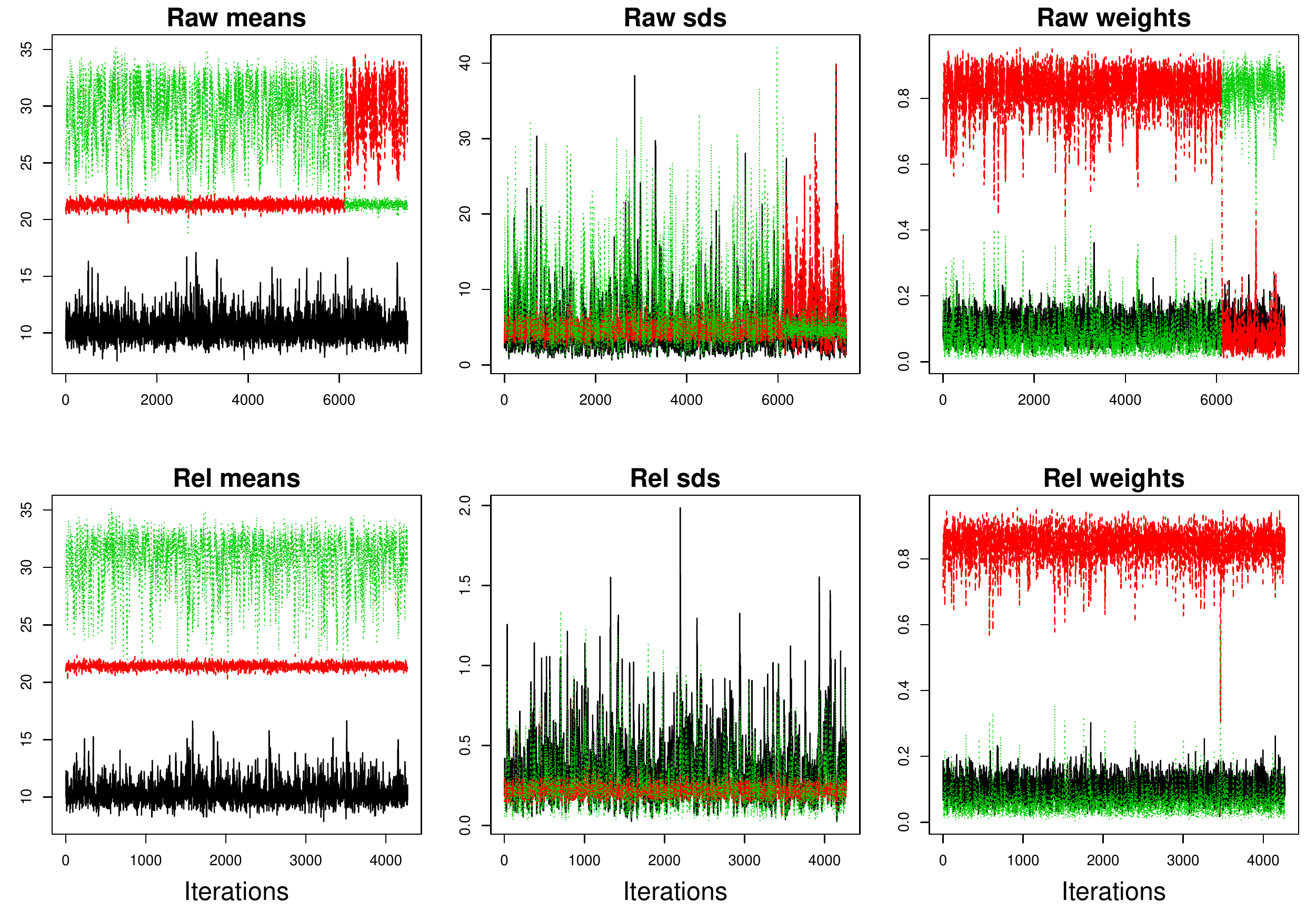}
\caption{Galaxies dataset: traceplots for the parameters $(\bm{\mu}, \bm{\sigma}, \bm{\eta})$, using the  \texttt{rjags} option (Gibbs sampling) with $k=3$, and $5000$ MCMC iterations. Top row: Raw MCMC outputs. Bottom row: Relabelled MCMC samples.}
\label{galaxies_chains}
\end{figure}
Figure~\ref{galaxies_chains} displays the raw chains and the relabelled chains for the parameter $\bm{\mu}$, $\bm{\sigma}$ and $\bm{\eta}$, respectively. As for the Fishery dataset, the pivotal relabelling algorithm permutes the means to undo the label switching phenomenon.

\section{Summary and discussion} 
\label{sec:summary}

The \texttt{pivmet} package proposes a variety of methods to identify pivotal units when a grouping structure is exhibited by the data. There are many statistical applications for which pivotal algorithms may be useful. In this paper, more emphasis is given to Bayesian mixture models whereas marginal attention has been devoted to robust $k$-means clustering and Dirichlet process mixtures.

Concerning the main application, the package performs a relabelling algorithm in order to deal with the problem of label switching in MCMC outputs for exploration of posteriors from mixture models. The input of the function is simply available from the MCMC output, and pivotal identification criterion used can be easily specified by the user. In addition, the package includes functions to fit a variety of Gaussian mixture models either by JAGS or Stan software. The relabelling method implemented by the \texttt{pivmet} package is computationally efficient, also compared to other available methods (for additional details, see \cite{egidi2018relabelling}). The computational overload is avoided by using a pivotal allocation step when relabelling, and by considering only a portion of the chains representing an approximation of the posterior conditional to being $k$ non-empty groups. For a small number of components ($k<5$)  all pivotal methods can be applied. When $k$ is large, the user should consider those available methods that directly maximize or minimize a given objective function from a co-association matrix obtained from the MCMC sample.  

The potential of pivotal methods for consensus clustering is also explored by the package, with application to the $k$-means clustering framework, where pivotal methods are useful tools for initializing the $k$-means procedure, to gain a robust solution.

The inclusion of additional functions allowing the estimation of the number of components, which is often unknown, either via posterior distribution for $k$ or predictive information criteria, is subject of ongoing work.


\end{document}